\documentclass{article}
\usepackage{spconf,amsmath,graphicx}
\usepackage[nocompress]{cite}
\usepackage{url}
\usepackage{hyperref}
\usepackage{amssymb,amsthm}
\usepackage{algorithm, algpseudocode}
\usepackage{booktabs}

\newtheorem{theorem}{Theorem}
\DeclareMathOperator*{\argmin}{arg\,min}

\def\H{{^\mathrm H}}
\def\bx{{\mathbf x}}
\def\bA{{\mathbf A}}
\def\by{{\mathbf y}}
\def\bU{{\mathbf U}}
\def\bF{{\mathbf F}}
\def\bS{{\mathbf S}}
\def\bn{{\mathbf n}}

\def\bD{{\mathbf D}}

\title{Breaking the Weak Recovery Limit in Random Phase Retrieval \\ with Learned Regularizers}

\name{Stanislas Ducotterd, Zhiyuan Hu, Michael Unser, and Jonathan Dong}
\address{Biomedical Imaging Group, \'Ecole Polytechnique F\'ed\'erale de Lausanne}

\begin{document}
\maketitle

\begin{abstract}
We seek to recover an unknown signal from nonlinear amplitude-only measurements, a challenging inverse problem. 
Strong theoretical guarantees have been established for idealized random measurements, defining the sampling ratio required for signal recovery. 
However, these results neglect signal priors, which can fundamentally shift these limits, potentially enabling reconstruction with far fewer measurements and simpler models.
We evaluate a variety of image priors in the context of severe undersampling with physically-grounded random measurement models. 
Our results show that these priors enable accurate recovery well below the weak recovery limit, the theoretical threshold required for recovery better than a random guess.
\end{abstract}


\section{Introduction}\label{sec:intro}

Phase retrieval addresses the problem of the recovery of a signal from intensity-only measurements, as physical detectors typically capture only the intensity of electric fields and not their phase \cite{dong2023phase}. 
Formally, we seek to recover a signal $\bx \in \mathbb{C}^n$ from measurements $\by \in \mathbb{R}^m$ given by

\begin{equation}\label{eq:first_eq}
    \by = |\bA \bx|,
\end{equation}
where the forward matrix $\bA \in \mathbb{C}^{m \times n}$ models the imaging system and $|\cdot|$ denotes the elementwise magnitude. 
The loss of phase information yields a nonconvex inverse problem. 

The phase retrieval problem was first investigated in the context of astronomy \cite{fienup1993hubble} and X-ray crystallography \cite{miao1999extending}. 
In this context, the forward matrix $\mathbf A$ is a discrete Fourier transform. 
Beyond the single Fourier transform setting, modern phase retrieval modalities rely on coded illumination strategies to acquire more measurements and facilitate the reconstruction process, as in ptychography~\cite{thibault2009probe,valzania2021accelerating} and Fourier ptychography~\cite{zheng2013wide,yeh2015experimental}. 
These modalities involve forward matrices composed of discrete Fourier transforms and diagonal modulation matrices, relying on the fast Fourier transform for efficient computation. 

To establish theoretical guarantees, one typically assumes that the forward matrix $\bA$ has i.i.d. Gaussian entries. 
In this setting, the sampling ratio $\alpha = m/n$ plays a central role: weak recovery (recovery better than random guess) is possible as soon as $\alpha >1$~\cite{mondelli2018fundamental}, while perfect recovery requires $\alpha>2$~\cite{maillard2020phase}. 
These thresholds on $\alpha$ determine the minimal number of measurements required for successful reconstruction. 

The assumption that $\bA$ is a Gaussian matrix has two major limitations: it does not capture the physical reality of most phase retrieval applications; and the dense matrix $\bA$ with $\mathcal{O}(n^2)$ entries is prohibitive due to both memory and computational complexity. 
The structured random framework introduced in \cite{hu2025structuredA, hu2025structured} leverages alternating discrete Fourier transforms and random diagonal matrices. 
It resolves the two issues as it is physically more realistic, admits a simple optical implementation with lenses and diffusers, and is faster to compute, with a computational cost of $\mathcal{O}(n \log n)$. 
Empirically, this architecture exactly matches the recovery thresholds of dense Gaussian matrices.

This framework provides an efficient and physically realistic forward model but remains limited by the same fundamental recovery thresholds. To bypass these limits, prior work has explored image priors for phase retrieval in other settings. Generative priors lower the recovery threshold~\cite{hand2018phase, bohra2023bayesian}, but have only been applied to real-valued, domain-specific datasets such as MNIST or CelebA where a generator can closely capture the data distribution. Image regularizers have also been extended to complex-valued phase retrieval~\cite{Isil, pnp_ptychography, bostan2020deep, sinha2017lensless}, but only in the oversampled regime, using oversampled Fourier measurements or multiple diffraction patterns from overlapping illuminations as in ptychography.

In this work, we adapt and compare a broad spectrum of image priors for complex-valued random phase retrieval in the heavily undersampled regime. We include classical total variation (TV)~\cite{rudin_nonlinear_1992} and the learned priors enabled by the plug-and-play (PnP) framework~\cite{6737048}, which inserts off-the-shelf neural denoisers~\cite{zhang2017beyond,zhang2018ffdnet,zhang_plug_and_play_2022}) into iterative reconstruction. We further evaluate the recent family of learned explicit regularizers~\cite{10223264,23M1565243,mfoe}. These regularizers learn rich data-driven priors while retaining a well-defined objective and convergence guarantees. 


Our technical contributions are the following. First, we extend the framework of learned explicit regularizers from real-valued linear inverse problems to complex-valued phase retrieval. Second, we derive a closed-form expression for the proximal operator of the amplitude-based data-fidelity term, enabling stable and efficient proximal optimization. Third, we show that image priors enable accurate recovery in the heavily undersampled regime, well below the weak recovery threshold $\alpha_\text{WR}=1$. Finally, we demonstrate that learned explicit regularizers, most notably based on multivariate fields of experts (MFoE)~\cite{mfoe}, outperform PnP with state-of-the-art denoisers such as DRUNet~\cite{zhang_plug_and_play_2022}. This suggests that principled explicit regularizers are especially well-suited to complex-valued nonlinear inverse problems like phase retrieval.

\section{Problem definition}
\label{sec:forward_operator}

Our aim is to recover a complex-valued signal 
$\bx \in \mathbb{C}^n$ from undersampled measurements 
$\by \in \mathbb{R}^m$ obtained through the nonlinear forward model  
\begin{equation}
    \by = \bigl|\bS\bU\bx\bigr| + \bn,
    \label{eq:phasemodel}
\end{equation}
where the main transform $\bU \in \mathbb{C}^{n \times n}$ is a unitary transform that models the imaging system, the subsampling mask $\bS \in \mathbb{C}^{m \times n}$ is a rectangular matrix composed of an identity matrix of size $(m \times m)$ and zeros elsewhere, and the additive noise vector $\bn \in \mathbb{C}^m$ is drawn from $\mathcal{N}(\mathbf 0, \sigma_{\bn}^2 \mathbf I)$. The undersampling case corresponds to $m \leq n$.

\begin{figure}
    \centering
    \includegraphics[width=\linewidth]{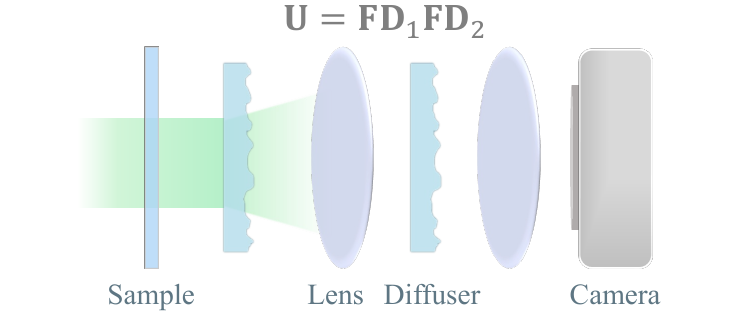}
    \vspace{-1em}
    \caption{Structured random model and its optical implementation using lenses (Fourier transforms) and diffusers (random diagonal matrices).}
    \label{fig:setup}
\end{figure}

In this work, we consider structured random models, introduced as a computationally efficient alternative to i.i.d. random matrices~\cite{hu2025structured}.  The unitary transform is given by a cascade of two Fourier and diagonal matrices,
\begin{equation}
    \mathbf U = \mathbf{FD}_1 \mathbf{FD}_2,
\end{equation}
where $\mathbf D_\ell$ is a random diagonal matrix with elements drawn i.i.d. uniformly in $\{-1,+1\}$.
The operators $\bD_\ell$ and $\bF$ accurately model the physical combination of diffusers and lenses in optical systems~\cite{dong2023phase}, as depicted in Figure~\ref{fig:setup}. 
It has been shown that this two-layer structure can emulate the robust properties of dense random matrices for phase retrieval \cite{hu2025structured}.

A key hyperparameter is the sampling ratio $\alpha = m/n$, which quantifies the ratio of the number of measurements to the number of unknowns. 
Small values of $\alpha$ correspond to a severe undersampling and make for a hard recovery problem.
Two thresholds characterize the feasibility of recovery in the high-dimensional limit $m, n \rightarrow \infty$. 
Below the weak recovery threshold $\alpha_\text{WR}$, no estimator can achieve a non-trivial correlation with $\bx$; partial recovery is only possible above it. For noiseless complex-valued phase retrieval with a Gaussian measurement matrix, $\alpha_\text{WR} = 1$ \cite{mondelli2018fundamental}. 
In this study, we focus on the regime $\alpha \leq 1$, where recovery is provably impossible without prior information on the image to recover.

\section{Methods}
\label{sec:methods}

Iterative algorithms are typically required to solve the nonlinear phase retrieval problem with image priors. We consider two main approaches: variational methods, which minimize a well-defined explicit objective, and plug-and-play methods, which incorporate learned denoisers into an iterative scheme.

\subsection{Variational Methods}

Variational methods solve the phase retrieval problem 
\begin{equation}\label{eq:main_obj}
    \min_{\bx \in \mathbb C^n} f(\bx) + \lambda \mathcal{R}_\sigma(\bx)
\end{equation}
by minimizing a well-defined loss function, composed of a data-fidelity term $f(\bx) = \tfrac{1}{2}\||\bS\bU\bx| - \by\|^2$ and a regularization term $\mathcal R_\sigma$, either TV or a learning-based explicit regularizers. The constant $\lambda$ is a hyperparameter that tunes the regularization strength.

TV promotes the sparsity of the finite differences of the image by penalizing the $\ell_2$-norm of the horizontal and vertical differences. Here, we instead use the Moreau envelope of the $\ell_2$-norm to allow for differentiability as in \cite{chambolle2016introduction}.

The convex ridge regularizer (CRR)~\cite{10223264}, weakly convex ridge regularizer (WCRR)~\cite{23M1565243}, and the multivariate fields of experts (MFoE)~\cite{mfoe} regularizers can be seen as learning-based evolution of TV. They propose to learn both the filters (which would correspond to the finite differences in TV) and the nonlinearity (which would correspond to the $\ell_2$-norm in TV) in order to maximize the reconstruction performance. Those regularizers can be described as
\begin{equation}
    R(\mathbf x) = \sum_{k=1}^K \langle \mathbf 1_n, \psi_k(\mathbf W_k \mathbf x)\rangle,
\end{equation}
where $\mathbf W_k$ denotes a convolution matrix and $\psi_k$ is a nonlinearity. CRR parameterizes the gradient of $\psi_k$ through a univariate nondecreasing linear spline in order to enforce the convexity of $\psi_k$. WCRR proposes to relax this approach by allowing $\psi_k$ to be nonconvex. Finally, MFoE generalizes both CRR and WCRR by utilizing multivariate potential functions, allowing the model to capture complex cross-filter interactions.

CRR, WCRR, and MFoE all learn a noise-dependent regularizer $R_\sigma \colon \mathbb{R}^n \to \mathbb{R}_+$ such that, for any image $\bx \in \mathbb R^n$ and $\by \sim \mathcal N(\bx,\sigma^2\mathbf I)$, we have $\operatorname{prox}_{R_\sigma}(\mathbf y) \approx \mathbf x$. The parameter $\sigma$ adapts the regularizer to the noise level.

All those regularizers were constructed with real-valued data in mind. To extend them to complex data, we regularize magnitude and normalized phase separately (ignoring phase wrapping), as in 
\begin{equation}
    \mathcal{R}_\sigma(\bx) = R_\sigma(|\bx|) + R_\sigma\!\left(\tfrac{\arg(\bx)}{2\pi}\right).
\end{equation} 

We solve \eqref{eq:main_obj} via accelerated proximal gradient descent with restart (Algorithm~\ref{alg:fast_restart}).
The algorithm follows a well-established FISTA-type acceleration scheme for faster convergence \cite{beck2009fast}. 
The restart mechanism ensures the robustness of the method despite the complex landscape that results from the nonlinearities of both the phase retrieval equation and the nonconvex regularizer. 
We set the stepsize $\gamma=1/( \operatorname{Lip}(\nabla R_\sigma))$. 

\begin{algorithm}[t]
\caption{Accelerated proximal gradient descent with restart}
\label{alg:fast_restart}
\begin{algorithmic}[1]
\Statex \textbf{Input:} Initialization $\bx_0 \in \mathbb{C}^n$, regularization strength $\lambda$, regularization noise level $\sigma$, and tolerance $\varepsilon > 0$
\State Set $t_0 = 1$, $k = 0$, $r_0 = \infty$, $s_0 = \infty$, $\mathbf{z}_0 = \bx_0$
\While{$r_k > \varepsilon$}
    \State $\bx_{k+1} = \operatorname{prox}_{\gamma f/ \lambda}\left(\mathbf{z}_{k} - \gamma  \nabla R_\sigma(\mathbf{z}_{k})\right)$
    \State $s_{k+1} = f(\mathbf z_k) + \lambda \mathcal{R}_\sigma(\mathbf z_k)$
    \State $t_{k+1} = \frac{1 + \sqrt{1 + 4t_{k}^2}}{2}$
    \State $\mathbf{z}_{k+1} = \bx_{k+1} + \frac{t_{k} - 1}{t_{k+1}} (\bx_{k+1} - \bx_{k})$
    \If{$s_{k+1} > s_{k}$}
        \State $\mathbf{z}_{k+1} = \bx_{k+1}$
        \State $t_{k+1} = 1$
    \EndIf
    \State $r_{k+1} = \|\bx_{k+1} - \bx_k\| / \|\bx_{k+1}\|$
    \State $k \gets k + 1$
\EndWhile
\Statex \textbf{Output:} Approximate solution $\bx_{k+1}$
\end{algorithmic}
\end{algorithm}

\subsection{Plug-and-Play Methods}\label{sec:pnp}

PnP approaches replace the explicit regularizer in an iterative reconstruction scheme with an off-the-shelf denoiser. 
Despite their simplicity and empirical success, PnP methods face significant theoretical hurdles regarding convergence. 
Because a generic neural-network-based denoiser $D_\sigma$ is rarely a proper proximal mapping of any scalar potential $R_\sigma$, the iterates are not guaranteed to converge to a stationary point of an objective functional. 
In contrast, explicit regularizers can easily be enforced to be convergent without sacrificing performance.

We follow \cite{zhang_plug_and_play_2022} and apply Algorithm~\ref{alg:pnp} with DnCNN, FFDNet, and DRUNet neural networks $D_\sigma$. To handle complex data, we split the signal into magnitude and phase, apply $D_\sigma$ to both, and recombine.

\begin{algorithm}[t]
\caption{Plug-and-play Algorithm}
\label{alg:pnp}
\begin{algorithmic}[1]
\Statex \textbf{Input:} Initialization $\bx_0 \in \mathbb{C}^n$, number $K$ of steps, regularization parameter $\lambda$, measurement noise level $\sigma_{\bn}$, noise schedule $\sigma_k$
\State Set $k = 0$
\While{$k < K$}
    \State $\beta_k = \lambda \sigma_{\bn}^2/\sigma_k^2$
    \State $\bx_{k+1} = \operatorname{prox}_{f/\beta_k}\left(D_{\sigma_k}(\bx_{k})\right)$
    
    \State $k \gets k + 1$
\EndWhile
\Statex \textbf{Output:} Approximate solution $\bx_{k+1}$
\end{algorithmic}
\end{algorithm}

\subsection{Data-Fitting Term and Its Proximal Operator}
\label{sec:proximal expression}

While gradient descent methods are often used for nonlinear optimization, proximal methods are significantly more stable. They are particularly attractive for phase retrieval since the amplitude-based loss~\cite{yeh2015experimental} admits a closed-form proximal operator. In practice, we have observed that proximal methods outperform gradient descent on this problem, offering faster convergence and being less sensitive to hyperparameter tuning. The proximal operator of the data-fidelity term is defined as
\begin{equation}
    \operatorname{prox}_{\lambda f}(\bx) = \argmin_{\mathbf z \in \mathbb C^n} \tfrac{1}{2}\|\mathbf z - \bx\|^2 + \tfrac{\lambda}{2} \||\bS\bU\mathbf z| - \mathbf y\|^2.
\end{equation}
In Theorem \ref{thm:prox}, we provide an exact closed-form expression for this proximal operator.

\begin{theorem}\label{thm:prox}
The proximal operator of the data-fitting term can be expressed as
\begin{align}\label{eq:prox}
    \operatorname{prox}_{\lambda f}(\bx) 
    &= \mathbf U\H \bS\left(\frac{1}{1 + \lambda}\mathbf{Ux} + \frac{\lambda}{1 + \lambda}\by \odot \mathrm{e}^{\mathrm j \arg \mathbf{(Ux)}}\right) \nonumber\\
   & \qquad + \mathbf U\H (\mathbf I - \mathbf S)\mathbf{Ux}.
\end{align}
\end{theorem}

\begin{proof}
    For the scalar problem,
    \begin{align}\label{eq:scalar_prox}
    \operatorname{prox}_{\tfrac{\lambda}{2}(|\cdot|-y)^2}(x) 
    &= \argmin_{z \in \mathbb C} \tfrac{1}{2}|z-x|^2 + \tfrac{\lambda}{2}(|z|-y)^2 \nonumber \\
    &= \argmin_{z \in \mathbb C} \tfrac{1}{2}|z|^2 + \tfrac{1}{2}|x|^2 + \tfrac{\lambda}{2}(|z|-y)^2 \nonumber\\
    & \qquad - |z||x|\cos(\arg z - \arg x)
    \end{align}
    and the only term depending on $\arg z$ is minimized by setting $\arg z = \arg x$. Substituting the phase, one can compute the optimal magnitude, which yields
    \begin{equation}
    \operatorname{prox}_{\frac{\lambda}{2}(|\cdot|-y)^2}(x) = \frac{1}{1 + \lambda}x + \frac{\lambda}{1 + \lambda}y\mathrm e^{\mathrm j \arg x}.
    \end{equation}
    The multidimensional case with a binary mask gives
    \begin{align}\label{eq:obj_binary}
        &\argmin_{\mathbf z \in \mathbb C^n} \frac{1}{2}\|\mathbf z - \bx\|^2 + \frac{\lambda}{2}\||\mathbf{Sz}| - \by\|^2 \nonumber \\
        &\qquad = \bS\left(\frac{1}{1 + \lambda}\bx + \frac{\lambda}{1 + \lambda}\by \odot \mathrm e^{\mathrm j \arg \bx}\right) + (\mathbf I - \mathbf S)\bx.
    \end{align}
    This follows from the separability of the objective; entries with $S_{kk}=1$ use the scalar proximal operator \eqref{eq:scalar_prox} while entries where $S_{kk}=0$ use the identity mapping. Finally~\cite{Combettes2011},
    \begin{equation}
    \operatorname{prox}_{g(\mathbf U \cdot)}(\bx) = \mathbf U^\H \operatorname{prox}_g(\mathbf{Ux}),
    \end{equation}
which yields \eqref{eq:prox}.
\end{proof}
%


\section{Results}

\subsection{Setting}


To construct complex-valued signals, we start from real-valued grayscale images, map pixel values in $[0,1]$ linearly to phases in $[-\pi/2,\pi/2]$, and assign unit magnitude.

For TV, which does not depend on $\sigma$, we solve \eqref{eq:main_obj} once starting from complex Gaussian noise $\mathbf x_0\sim\mathcal{CN}(\mathbf 0,\mathbf I)$ using Algorithm \ref{alg:fast_restart}. For CRR, WCRR and MFoE, we exploit the noise parameter through a continuation scheme: we solve the problem three times, starting with $\sigma=1/4$, then $\sigma=1/16$, and finally $\sigma=1/64$ in the noiseless case. In the noisy case, the same strategy is applied, but starting at $\sigma=1$. The first run is initialized with $\mathbf x_0 \sim \mathcal{CN}(\mathbf 0,\mathbf I)$, subsequent runs with the previous solution. This progressively refines the reconstruction from coarse to fine.
Additionally, in the first run, we modify the data-fitting term to avoid poor local minima---whenever $\alpha>0.25$, we randomly subsample $\mathbf S$ to one quarter of its entries. 

Other hyperparameters include the relative tolerance for the stopping criterion $\varepsilon=10^{-5}$. To ensure a regularization effect that is robust with respect to the noise level, the regularization parameter in Algorithm \ref{alg:fast_restart} is scaled as $\lambda = \lambda_0 \sigma_{\bn}$, with baseline weights $\lambda_0 = \mathrm e^7$ for TV, $\lambda_0 = \mathrm e^5$ for CRR and WCRR, and $\lambda_0 = \mathrm e^9$ for MFoE. In the noiseless setting, the objective is evaluated in the constrained limit: the data-fitting term $f$ acts as the indicator function of the measurements, causing the proximal step in \eqref{eq:prox} to reduce exactly to a projection, while the restart logic tracks the regularization functional alone.

For the PnP experiment, the algorithm is initialized with $\mathbf x_0 \sim \mathcal{CN}(\mathbf 0,\mathbf I)$ and uses $\lambda=0.23$, as recommended in \cite{zhang_plug_and_play_2022}. The DnCNN is only trained on a fixed noise level (either 15/255, 25/255, or 50/255); we then use the model trained on 25/255, which stays fixed during the whole process.
For FFDNet and DRUNet, the noise schedule starts at $\sigma_0=1$ and decreases geometrically to $\sigma_K=1/(10\alpha)$ for FFDNet and $\sigma_K=1/(1000\alpha)$ for DRUNet, where $\alpha$ is the sampling ratio. 
We do not apply the data-fitting modification strategy of the explicit regularizers for the implicit ones as it does not result in performance improvements.

The hyperparameters of Algorithms~\ref{alg:fast_restart} and \ref{alg:pnp} were tuned on the Set12 dataset \cite{7839189}, while performance is reported on BSD68 \cite{937655}. All metrics correspond to averages over the test dataset.

\subsection{Visual and Quantitative Comparison}



\begin{figure*}[htbp] 
    \centering
    
    \includegraphics[width=0.9\textwidth]{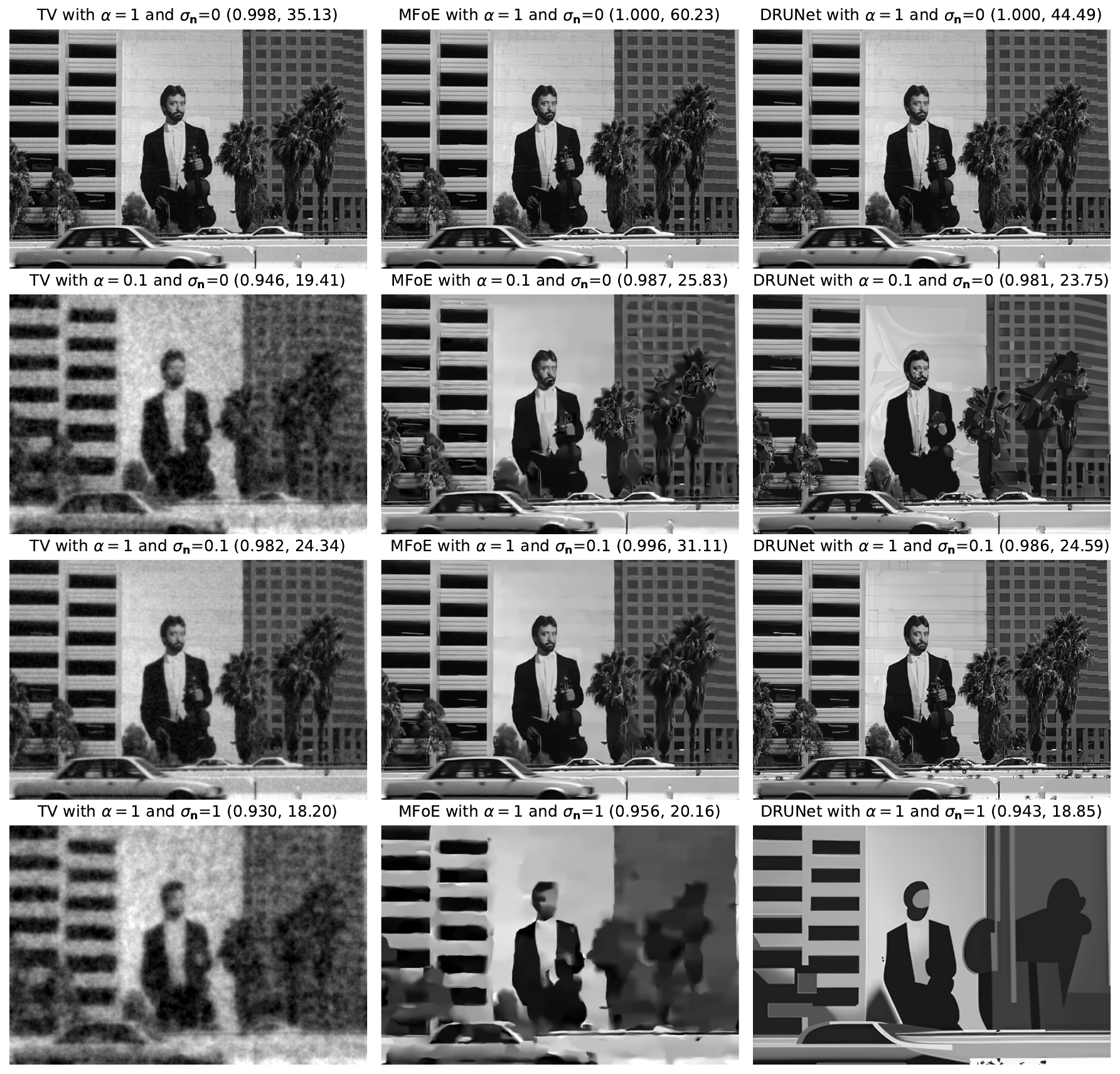}
    \vspace{-0.5em}
    \caption{Reconstruction results from TV, MFoE, and DRUNet. The reported metrics are (cosine similarity, PSNR).}
    \label{fig:results}
    
    \vspace{1.5em} 
    
    \begin{minipage}{0.48\textwidth}
        \centering
        \includegraphics[width=0.9\linewidth]{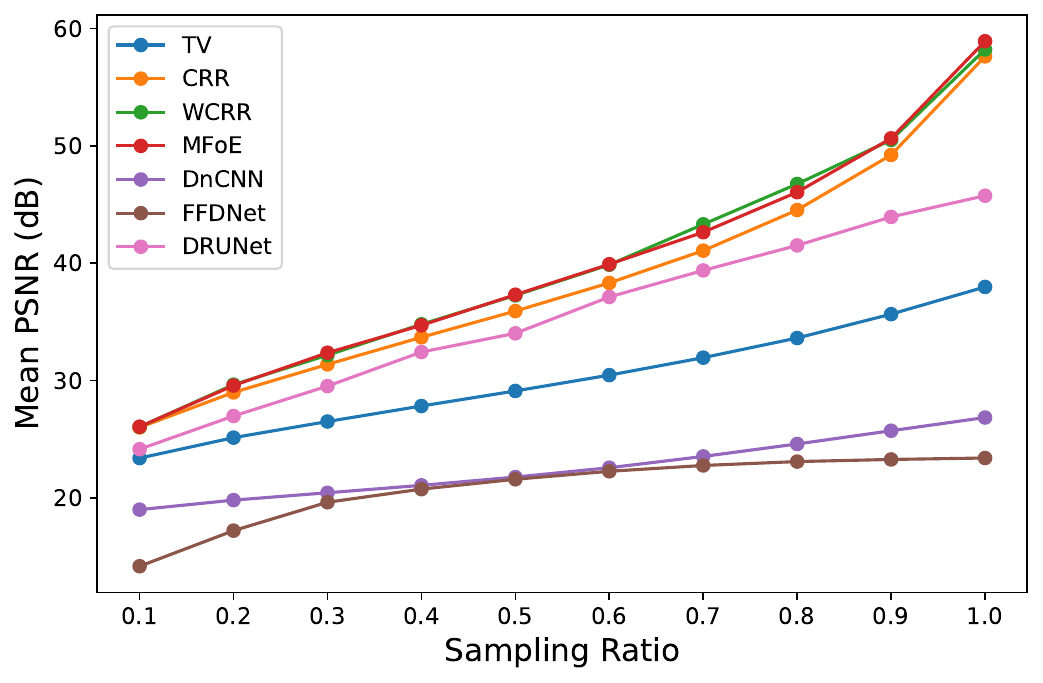}
        \vspace{-1em} 
        \caption{PSNR (dB) for each model and sampling ratio without noise.}
        \label{fig:alpha_psnr}
    \end{minipage}\hfill
    \begin{minipage}{0.48\textwidth}
        \centering
        \includegraphics[width=0.9\linewidth]{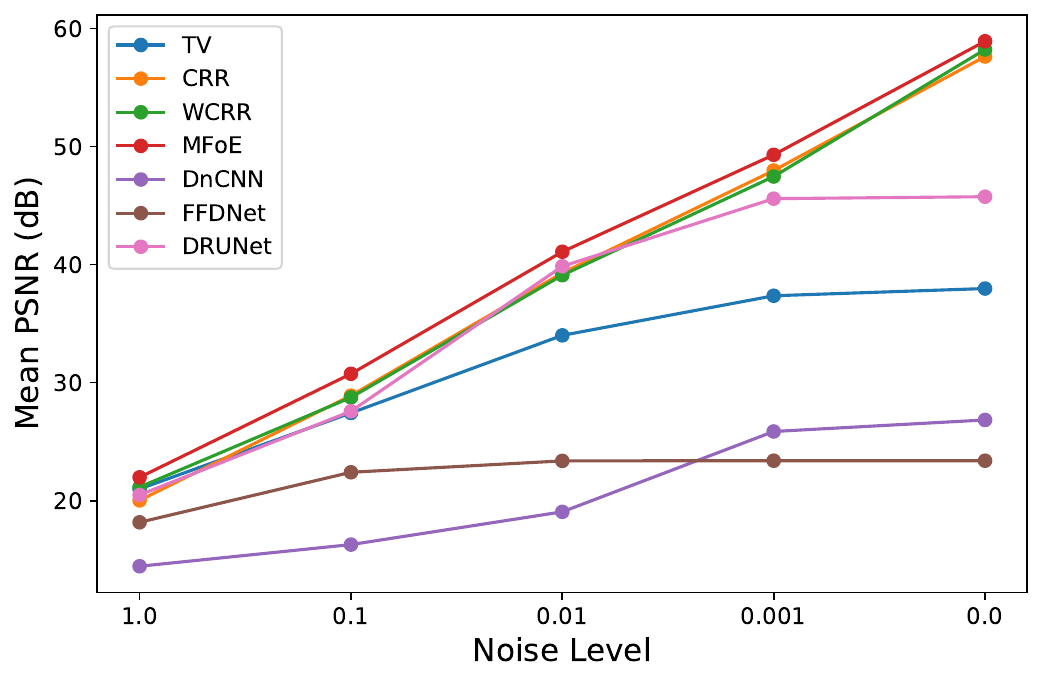}
        \vspace{-1em} 
        \caption{PSNR (dB) for each model and noise level without subsampling.}
        \label{fig:noise_psnr}
    \end{minipage}
\end{figure*}

Visual reconstructions are displayed in Figure~\ref{fig:results}. 
The reconstruction performance for the noiseless subsampled and noisy case are reported in Figure~\ref{fig:alpha_psnr} and \ref{fig:noise_psnr} respectively.
The metrics used to assess the performance of the reconstruction is the widely used peak signal-to-noise ratio (PSNR, higher is better). Interestingly, despite its state-of-the-art performance in denoising and multiple linear inverse problems, DRUNet is outperformed by CRR, WCRR and MFoE in most settings. This suggests that the PnP framework may be suboptimal in the context of challenging complex-valued nonlinear inverse problems. On top of that, MFoE and WCRR dominate the other priors in the noiseless setting, and MFoE stands out as the single best regularizer in the noisy setting.

For reference, plain gradient descent across all scenarios yields only about 4 dB of PSNR, which is not better than a random guess. This highlights the importance of image priors. Even a simple prior such as TV significantly boosts performance.



\section{Conclusion}

We investigated the benefit of image priors in challenging phase retrieval problems with noise or extreme undersampling. We considered explicit regularizers (TV, CRR, WCRR, and MFoE), and implicit deep-learning-based priors (DnCNN, FFFDNet, and DRUNet). Our experimental results show that the incorporation of priors significantly improves the reconstructions, even way below the weak recovery threshold. The explicit regularizers generally outperform the deep-learning-based ones, which highlights the effectiveness of learned explicit regularizers in the context of challenging nonlinear inverse problems.

\section{Acknowledgments}
Stanislas Ducotterd acknowledges funding from the European Research Council (Grant Agreement - Project No 101020573 FunLearn). Zhiyuan Hu and Jonathan Dong acknowledge funding from the Swiss National Science Foundation (Grant PZ00P2\_216211).

\begingroup
\small
\bibliographystyle{IEEEbib}
\bibliography{refs}
\endgroup

\end{document}